\documentclass[prl, aps, twocolumn, superscriptaddress]{revtex4-1}
\usepackage{graphicx,graphics,psfrag,amsmath,calc,mathtools}
\usepackage{amsfonts}
\usepackage{epsfig, bm}
\usepackage{color,comment}
\usepackage{cancel}
\usepackage{float}
\def\ba{\begin{eqnarray}}
\def\ea{\end{eqnarray}}

\usepackage{tikz,fp}
\usepackage{tikz-cd}
\usetikzlibrary{arrows}
\usetikzlibrary{intersections}
\usetikzlibrary{shapes.geometric}
\usetikzlibrary{decorations.pathmorphing, patterns,shapes,fixedpointarithmetic}
\usetikzlibrary{decorations.markings}
\pgfdeclarepatternformonly{south west lines}{\pgfqpoint{-0pt}{-0pt}}{\pgfqpoint{3pt}{3pt}}{\pgfqpoint{3pt}{3pt}}{
	\pgfsetlinewidth{0.4pt}
	\pgfpathmoveto{\pgfqpoint{0pt}{0pt}}
	\pgfpathlineto{\pgfqpoint{3pt}{3pt}}
	\pgfpathmoveto{\pgfqpoint{2.8pt}{-.2pt}}
	\pgfpathlineto{\pgfqpoint{3.2pt}{.2pt}}
	\pgfpathmoveto{\pgfqpoint{-.2pt}{2.8pt}}
	\pgfpathlineto{\pgfqpoint{.2pt}{3.2pt}}
	\pgfusepath{stroke}}

\tikzset{
	mid arrow/.style={postaction={decorate,decoration={
				markings,
				mark=at position .575 with {\arrow{stealth}}
	}}},
	near arrow/.style={postaction={decorate,decoration={
				markings,
				mark=at position .275 with {\arrow{stealth}}
	}}},
	far arrow/.style={postaction={decorate,decoration={
				markings,
				mark=at position .800 with {\arrow{stealth}}
	}}},
	snake arrow/.style={fixed point arithmetic, decorate, decoration={snake,amplitude=2pt, segment length=11pt},postaction={decoration={markings,mark=at position 0.625 with {\arrow{stealth}}},decorate}},
}

\usepackage{color}

\usetikzlibrary {decorations.fractals}
\usetikzlibrary { decorations.pathmorphing, decorations.pathreplacing, decorations.shapes, }

\begin{document}
\date{\today}

\title{Superradiant Transition to a Fermionic Quasicrystal in a Cavity}

\author{Bo-Hao Wu}
\affiliation{Department of Physics and Beijing Key Laboratory of Opto-electronic Functional Materials and Micro-nano Devices, Renmin University of China, Beijing 100872, China}
\affiliation{Key Laboratory of Quantum State Construction and Manipulation (Ministry of Education), Renmin University of China, Beijing 100872,China}
\author{Xin-Xin Yang}
\affiliation{Department of Physics and Beijing Key Laboratory of Opto-electronic Functional Materials and Micro-nano Devices, Renmin University of China, Beijing 100872, China}
\affiliation{Key Laboratory of Quantum State Construction and Manipulation (Ministry of Education), Renmin University of China, Beijing 100872,China}
\author{Wei Zhang}
\email{wzhangl@ruc.edu.cn}
\affiliation{Department of Physics and Beijing Key Laboratory of Opto-electronic Functional Materials and Micro-nano Devices, Renmin University of China, Beijing 100872, China}
\affiliation{Key Laboratory of Quantum State Construction and Manipulation (Ministry of Education), Renmin University of China, Beijing 100872,China}
\affiliation{Beijing Academy of Quantum Information Sciences, Beijing 100193, China}
\author{Yu Chen}
\email{ychen@gscaep.ac.cn}
\affiliation{Graduate School of China Academy of Engineering Physics, Beijing 100193, China}
\date{\today}

\begin{abstract}
Recently, the steady state superradiance in degenerate Fermi gases has been realized in a cavity, following the previous discovery of the Dicke transition in Bose gases. The most prominent signature of fermionic Dicke transition is its density dependence, which is manifested as the Fermi surface nesting effect and the Pauli blocking effect. We study the superradiant transition in one-dimensional Fermi gases in a cavity with the presence of an incommensurate dipolar lattice. We find a first-order Dicke transition induced by indirect resonance effect, which is a resonance between two atomic levels by the level repulsion from a third level, and causes extra gap opening. By formulating a phenomenological theory, we find that the critical pumping strength for this first-order Dicke transition shows a linear V-shape kink near a particular indirect resonance modified filling $\nu_{\rm IRM}$. The presence and the unique density dependence of this transition manifest the fermionic nature and verify the mechanism of the quasicrystal superradiant transition.
\end{abstract}

\maketitle

{\emph {Introduction.}}-- Recent developments in achieving strong coupling for atoms in a cavity~\cite{Esslinger_2007,Esslinger_2013} have enabled us to realize the Dicke transition in ultracold atomic gases~\cite{Esslinger_2010}. In Bose gases, a steady state superradiance accompanied with a self-organized checkerboard density wave has been observed~\cite{Esslinger_2011}, with the characteristic density order being verified by roton softening~\cite{Esslinger_2012} and critical exponent measurements for the dynamical structure factor~\cite{Esslinger_2015}. Lately, the long-desired fermionic steady state superradiance~\cite{Simons_2014,Piazza_2014,Chen_2014} has been realized, and the Pauli blocking effect has been witnessed in the high-density regime~\cite{Wu_2021}. The statistical effects in fermionic superradiance have then triggered a lot of interests~\cite{Kjaegaard_2021,Ketterle_2021,Ye_2021}.

Another intriguing scenario is the superradiant transition which leads to an incommensurate density order with the background lattice. The spontaneously emerged density order acts as a disordered potential, and the atoms must decide the extent of disorder in response to their own dynamics. In this quasicrystal (QC) superradiant phase, the single-particle dynamics are studied, and a L\'{e}vy walk is found~\cite{Zheng_Cooper}. For many-body phases, previous works predict that Anderson localization can help superradiance for Bose gases, resulting in a first-order Dicke transition~\cite{Sun_2020, Piazza_2019}.
\begin{figure}[t]
\includegraphics[width=8.2cm]{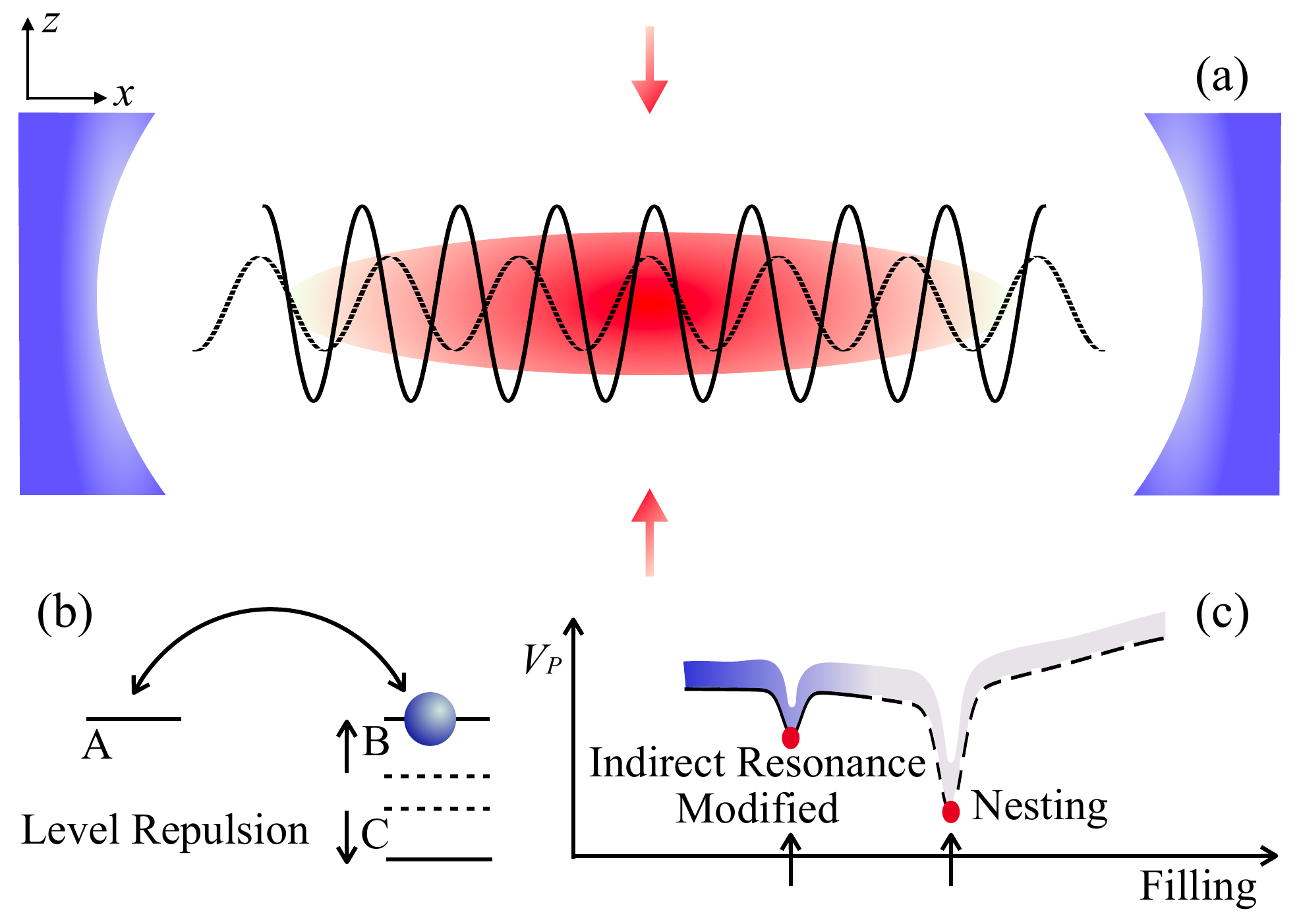}
\caption{(a) A 1D Fermi gas placed in a cavity with an incommensurate dipolar lattice along the $\hat{x}$ direction, and driven by pumping field along the $\hat{z}$ direction. (b) Mechanism of IR effect, a resonance by level repulsion. (c) Schematic phase diagram of the QC superradiant transition showing the critical pumping strength $V_p$ as a function of filling. The solid and dashed lines denote first-order and second-order transitions, respectively. The color above the line denotes the mean cavity photon number after transition with dark (light) color representing large (small) photon number. The Fermi surface nesting and IR effects are shown as dips.
}
\label{Mechanism}
\end{figure}

In this Letter, we study the superradiant transition of a one-dimensional (1D) Fermi gas in a cavity with an incommensurate dipolar lattice, as depicted in Fig.~\ref{Mechanism}(a). We find a density-dependent first-order QC superradiant transition induced by an indirect resonance (IR), across which the occupation energy changes abruptly owing to the modification of density of states (DOS) by extra gap opening. The mechanism for indirect resonance is similar to the scheme of variable-range hopping in interacting disordered systems~\cite{Mott56,Mott69,Mott79}, where two off-resonant localized states can be brought into resonance by the absorption of a phonon. Here for atoms in a cavity, a resonance can also take place between two states (A and B) as one of them (B) is shifted in energy by the level repulsion from a third level (C), assisted by the emergent superradiant field. This IR effect, as illustrated in Fig.~\ref{Mechanism}(b), will effectively open a gap at finite cavity condensation for a specific filling. The so-called IR gap will modify other gaps and consequently alter the dependence of occupation energy on the cavity field, such that the superradiant transition becomes strongly density dependent. The resulting phase diagram is schematically depicted in Fig.~\ref{Mechanism}(c). While the Fermi surface nesting effect leads to a vanishing critical pumping strength at the filling of perfect nesting, the participation of IR effect can cause a dip at a different filling, named indirect resonance modified (IRM) filling. The kink of critical pumping around such filling is a smoking-gun evidence of IR effect, which is explained by a phenomenological theory and verified by numerical simulation using experimentally relevant parameters.

{\emph {Model and Mean Field Theory.}}--
We consider a 1D spinless Fermi gas placed in a high finesse cavity. The Fermi gas is aligned along the $\hat{x}$ direction, which is also the cavity direction as shown in Fig.~\ref{Mechanism}(a). The system is subjected by two counter-propagating dipolar laser beams along the $\hat{x}$ direction, polarized in the $\hat{y}$ direction, and two pumping lasers along the $\hat{z}$ direction, polarized in the $\hat{y}$ direction. Both the pumping frequency and the cavity frequency are far detuned from the atomic excitation energy. By using a standard rotating-wave approximation and eliminating the electronic motion, we obtain the Hamiltonian ($\hbar=k_B =1$)
\ba
&&\hat{H} = \int \!dx \hat{\Psi}^{\dagger}(x) \hat{H}_0 \hat{\Psi}(x) - \Delta_c\hat{a}^{\dagger}\hat{a}, \\
&&\hat{H}_0 = \hat{H}_{\rm at}+\eta(x)(\hat{a}^{\dagger}+\hat{a})+U(x)\hat{a}^{\dagger}\hat{a},  \\
&&\hat{H}_{\rm at} = -\frac{\partial_x^2}{2m}-\mu+V_d(x),
\ea
where $\hat{\Psi}(x)$ is the field operator of fermions with chemical potential $\mu$, $\hat{a}$ is the cavity field operator, and $\Delta_c$ is the cavity field detuning. The dipolar potential is $V_d(x)=V_d\cos^2(k_d x)$ and the cavity field self-energy potential is $U(x)=U_0\cos^2 (k_c x)$, with corresponding wave numbers $k_d$ and $k_c$. The depth of $U_0=g_0^2/\Delta_{\rm AC}$ is determined by the atom--cavity coupling strength $g_0$ and the AC Stark shift $\Delta_{\rm AC}$. The interference lattice between the pumping field and the cavity field is $\eta(x)=\eta_0\cos(k_c x)$, where $\eta_0=\Omega_p g_0/\Delta_{\rm AC}$ with $\Omega_p$ the Rabi frequency of the pumping field. In the following discussion, we define the recoil energy $E_R=k_d^2/2m$ as the energy unit and choose $k_c/k_d = 2(\sqrt{2}-1)$ as a particular example.

We adopt the mean field approach and assume that $\alpha\equiv\langle\hat{a}\rangle$ at the steady state, where $\langle\cdot\rangle \equiv {\rm tr}(\cdot\hat{\rho}_{\rm st})$ represents an ensemble average over the steady state density matrix $\hat{\rho}_{\rm st}=e^{-\beta\hat{\cal H}_{0,\alpha}}/\cal{Z}_{\alpha}\otimes|\alpha\rangle\langle\alpha|$ with $\beta=1/T$ the inverse temperature. Here, $|\alpha\rangle$ is the coherent state of the cavity photon, ${\cal Z}_{\alpha}={\rm Tr} (e^{-\beta\hat{\cal H}_{0,\alpha}})$, and $\hat{\cal H}_{0,\alpha}=\langle\alpha|\int dx\hat{\Psi}^\dag(x)\hat{H}_0\hat{\Psi}(x)|\alpha\rangle$. This assumption is justified more comprehensively by the Keldysh Green's function method~\cite{Piazza_2014_2}. The dynamical equation for the cavity field then follows the steady equation $i\partial_t\alpha=\langle [\hat{a},\hat{H}]\rangle-i\kappa \alpha$, which can be written as 
\begin{eqnarray}
i\partial_t\alpha=\partial_{\alpha^*}{\cal F}_\alpha-i\kappa\alpha.
\end{eqnarray}
Here, $\kappa$ is the cavity decay rate and the free energy is given by ${\cal F}_\alpha\equiv -T\log {\cal Z}_{\alpha}-\Delta_c\alpha^*\alpha$. One can check $-T\partial \log{\cal Z}_\alpha/\partial \alpha^*=\eta_0\Theta+U_0 {\cal B}\alpha$, where $\Theta={\rm Tr}(\hat{\Theta}e^{-\beta\hat{\cal H}_{0,\alpha}})/{\cal Z}_\alpha$ is the fermion density order with $\hat{\Theta}=\int dx \hat{\Psi}^\dag(x)\eta(x)\hat{\Psi}(x)/\eta_0$, and ${\cal B}={\rm Tr}(\hat{\cal B}e^{-\beta\hat{\cal H}_{0,\alpha}})/{\cal Z}_\alpha$ is another density order with $\hat{\cal B}\equiv \int dx \hat{\Psi}^\dag(x)U(x)\hat{\Psi}(x)/U_0$. For simplicity, we assume that the temperature of atoms is zero, and ${\cal F}_\alpha$ is just the ground state energy $E_\alpha$.

For a steady state, we have $\partial_t\alpha=0$, that translates to
\begin{equation}
\alpha = \frac{\eta_0\Theta}{\tilde{\Delta}_c+i\kappa},
\label{Steady}
\end{equation}
with $\tilde{\Delta}_c\equiv\Delta_c-U_0{\cal B}\approx \Delta_c-U_0N_{\rm at}/2$ and $N_{\rm at}$ the atom number. This steady state equation can in principle be solved numerically, but the outcome may be difficult to interpret to identify the underlying physics. For instead, we adopt an equivalent method to minimize the energy $E_\alpha$. First of all, we notice that the density order $\Theta$ is real. Therefore, the phase of the steady state cavity field is determined by the ratio of $\tilde{\Delta}_c$ and $\kappa$. Secondly, although there is another density order ${\cal B}$ controlled by the $U_0$ term and $\tilde{\Delta}_c$ is in general not a constant, for simplicity we can assume a large $\tilde{\Delta}_c$ such that the phase of the cavity field is approximately a constant. Under this condition, the minimization of $E_\alpha$ with respect to $\alpha$ is equivalent to solving the steady state equation.

To obtain the ground state energy of the atomic field for a virtually condensed $\alpha$, we need to calculate the eigenvalues of $\hat{H}_0(\alpha)=\langle \alpha|\hat{H}_0|\alpha\rangle$. By denoting the eigenvalue of $\hat{H}_0(\alpha)$ as $\varepsilon_{n,\alpha}$, such that $\hat{H}_0(\alpha)\phi_{n,\alpha}=\varepsilon_{n,\alpha}\phi_{n,\alpha}$, we obtain the ground state energy
\begin{equation}
E_\alpha=\sum_n\varepsilon_{n,\alpha} \theta(\varepsilon_F-\varepsilon_{n,\alpha}) - \Delta_c\alpha^*\alpha,\label{MFT}
\end{equation}
where $\theta(x)$ is the Heaviside step function and $\varepsilon_F$ is the Fermi energy fixed by the filling fraction defined in the thermodynamic limit. The expression above is composed by two contributions, including the vacuum energy $E_{\rm vac}=\varepsilon_{0,\alpha}N_{at}-\Delta_c\alpha^*\alpha$ and the occupation energy $E_{\rm occ}=\sum_{n}(\varepsilon_{n,\alpha}-\varepsilon_{0,\alpha}) \theta(\varepsilon_F-\varepsilon_{n,\alpha})$. Fermi statistics and the consequent density-dependent effect are mainly manifested in $E_{\rm occ}$.

{\emph {Nesting and Indirect Resonance Effects}}.--
Next we present the results of minimization of $E_\alpha$ for the 1D Fermi gas under open boundary conditions. We first show some numerical results, then introduce an effective theory to explain the nesting and IR effects. For numerical treatment, we first solve the Bloch states $|k\rangle$ as the eigenstates of $\hat{H}_{\rm at}$ with only the dipolar lattice $V_d(x)$. By focusing on the lowest band, we construct the Wannier basis $|j\rangle=\sum_{k}e^{i kj}|k\rangle$ using Bloch states $|k\rangle$ with $k \in[-k_d,k_d)$ in the first Brillouin zone. The Hamiltonian $\hat{H}_0(\alpha)$ can be expanded in the basis of $|j\rangle$ with a finite-size cutoff and then solved by numerical diagonalization.

\begin{figure}[t]
\includegraphics[width=8.cm]{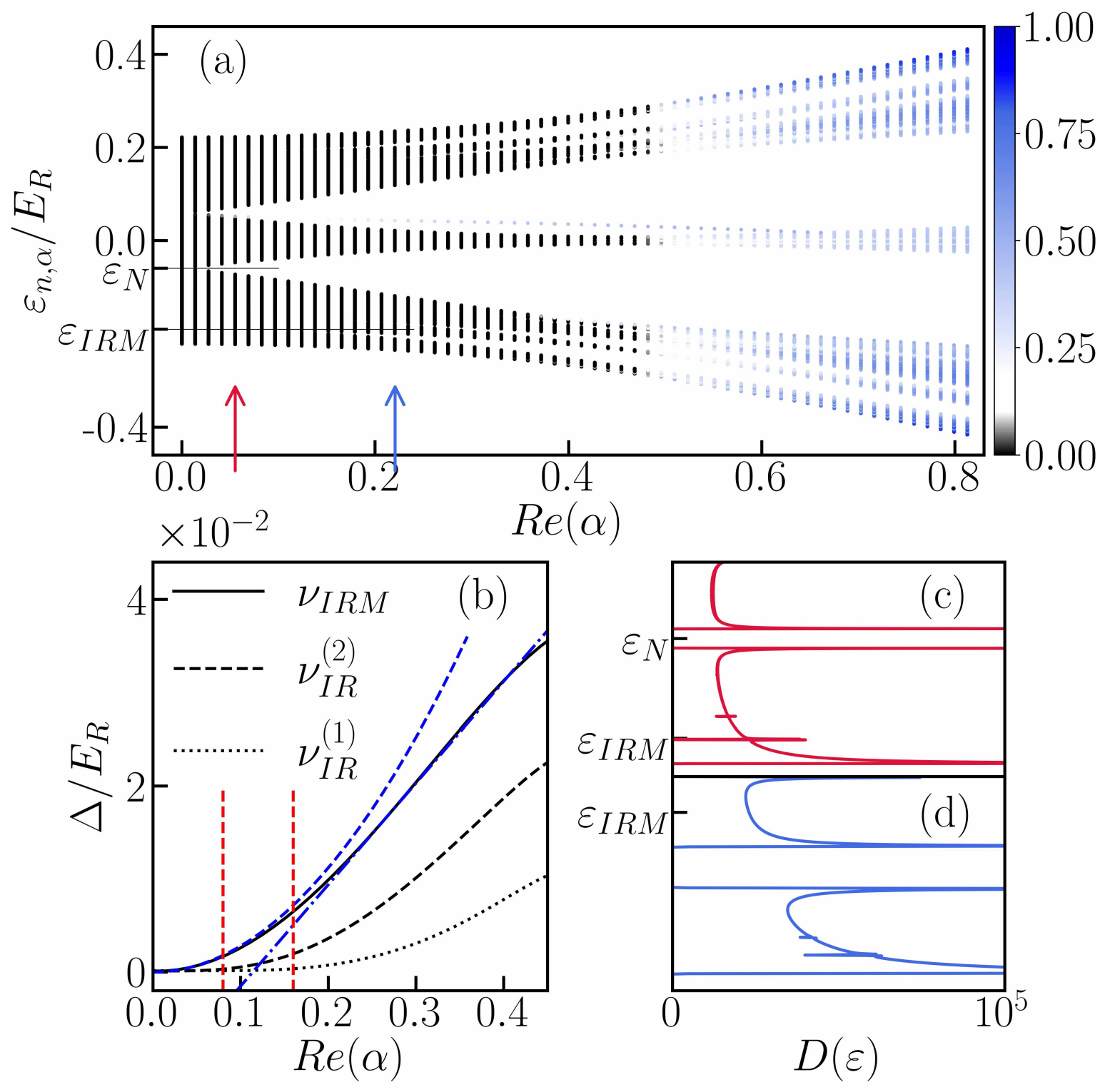}
\caption{(a) The single-particle spectrum of $\hat{H}_0$ as a function of Re($\alpha$) and the IPR of all eigenstates (false color). (b) The gap widths $\Delta/E_R$ as a function of Re($\alpha$) at different fillings $\nu_{\rm IRM}=1-k_c/k_d$, $\nu^{(1)}_{\rm IR}=2-2k_c/k_d$, and $\nu^{(2)}_{\rm IR}=3k_c/2k_d-1$. The red dashed lines correspond to IR gap opened points. The blue dashed and dash-dotted lines are quadratic and linear fit of the dispersion of $\nu_{\rm IRM}$, respectively. (c,d) The DOS around nesting and IRM fillings with different Re($\alpha$), labeled by red and blue arrows in (a). Here $V_d/E_R=3$, $\eta_0/E_R=-0.2733$, $U_0/E_R=-0.01$, $\Delta_c/E_RN_{\rm at}=-0.155$, $\kappa/E_RN_{\rm at}=0.0075$ and $N=4000$ ($N$ is the system size).
}
\label{Spectrum}
\end{figure}

In Fig.~\ref{Spectrum}(a), we show the single-particle spectrum of $\hat{H}_0$ as a function of Re$(\alpha)$, which characterizes the depth of the superradiant lattice. The results present several apparent gaps at different fillings (thin solid lines). Firstly, one can see a gap opening at zero cavity field $\alpha=0$ at a special energy $\varepsilon_{\rm N}$. The filling of this energy is $\nu_{\rm N} = k_c/2k_d$, at which the states $|\pm k_c/2\rangle$ are resonant. This is addressed as a Fermi surface nesting effect, which leads to a direct resonance gap of width being proportional to $|\alpha|$. Besides, other gaps can be also seen at filling factors $\nu_{\rm IRM}=1-k_c/k_d$, $\nu^{(1)}_{\rm IR}=2-2k_c/k_d$, and $\nu^{(2)}_{\rm IR}=3k_c/2k_d-1$, and the gap widths as functions of Re$(\alpha)$ are given in Fig.~\ref{Spectrum}(b).

For the gaps at $\nu^{(1)}_{\rm IR}=2-2k_c/k_d$ and $\nu^{(2)}_{\rm IR}=3k_c/2k_d-1$, one can clearly find that these gaps are opened at finite $\alpha$ [red dashed lines in Fig.~\ref{Spectrum}(b)], hence are classified as IR gaps. Take $\nu^{(1)}_{\rm IR}$ as an example. Originally, $|2k_d-2k_c\rangle$ and $|2k_d-3k_c\rangle$ are close to each other but not resonant. The $\eta$ term gives a strong repulsion between these two levels and one level becomes resonant with $|k_d\rangle$, causing an IR gap opening. Similar phenomenon is also observed for filling $\nu^{(2)}_{\rm IR}=3k_c/2k_d-1$. In addition, the new opened IR gaps can modify the existing direct resonance gaps as a result of level repulsion, that is, every new gap will compress the existing gaps. Specifically, at the filling $\nu_{\rm IRM}=1-k_c/k_d$, a gap is present for arbitrarily small $\alpha=0$ at an energy $\varepsilon_{\rm IRM}$, owing to the direct resonance process with an intensity $U_0$ and a $2k_c$ momentum transfer. Thus, this gap should be proportional to $|\alpha|^2$ [blue dashed line in Fig.~\ref{Spectrum}(b)], at least for small $|\alpha|$. However, when an IR gap opens at a finite $|\alpha|$, the level repulsion effect will alter the $|\alpha|^2$ behavior of the direct resonance gaps, and causes a linear dependence [blue dash-dotted line in Fig.~\ref{Spectrum}(b)] on the cavity condensation strength. 

\begin{figure}[t]
\includegraphics[width=7.5cm]{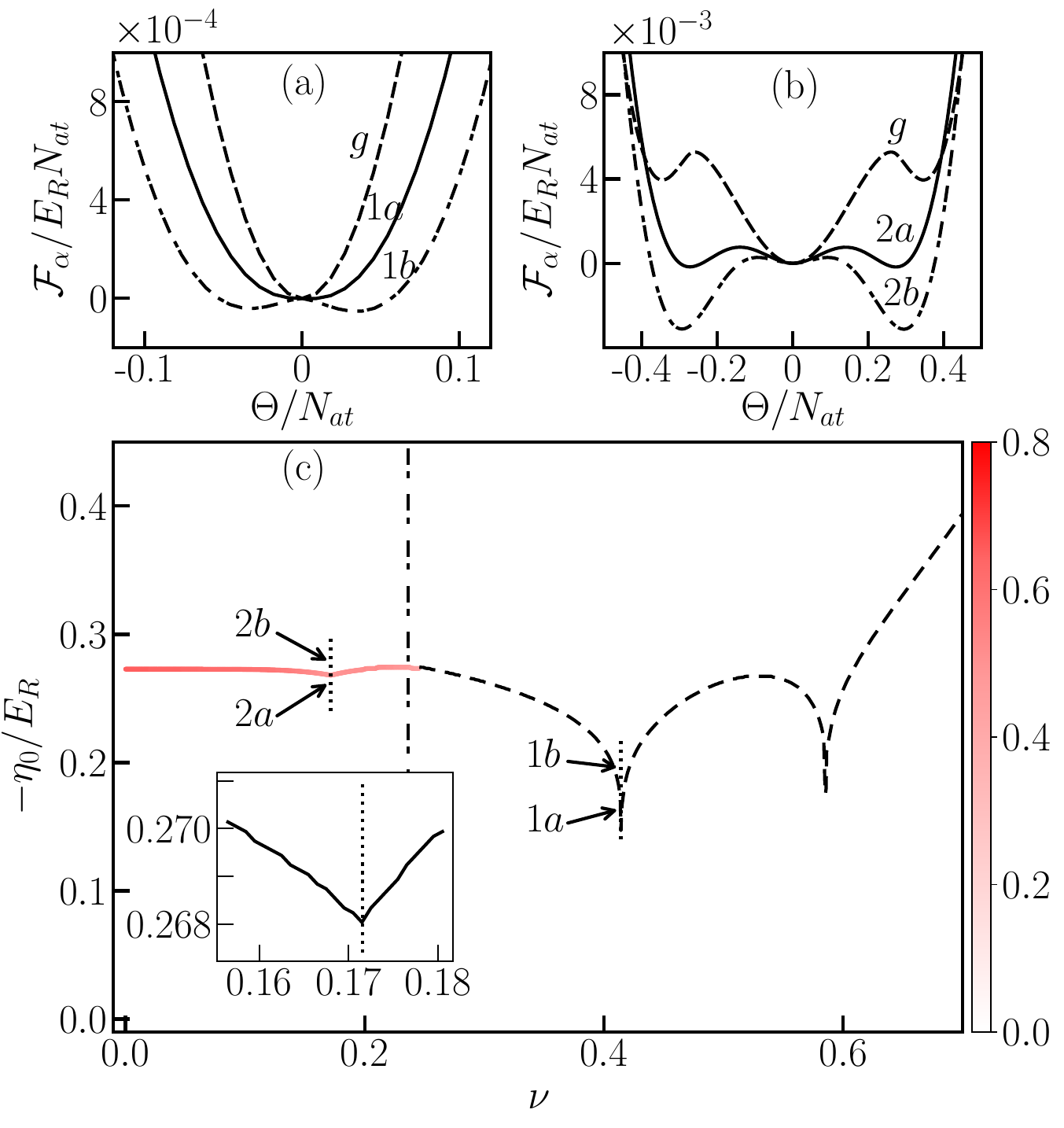}
\caption{(a,b) The ground state energy density ${\cal F}_{\alpha}/E_R N_{\rm at}$ as a function of density order $\Theta/N_{\rm at}$ at positions marked as 1a, 1b, 2a, 2b in (c). The dashed $g$ lines denote the vacuum energy density $E_{\rm vac}/E_RN_{\rm at}$ at phase boundary. (c) The phase diagram in terms of filling factor $\nu$ and critical pumping strength $\eta_0/E_R$. The solid (dashed) line represents the first-order (second-order) phase boundary and the false color at first-order phase boundary denotes the cavity field $|\alpha|$. The dash-dotted line denotes the boundary between Anderson localization phase and quasi-crystal phase. The inset shows the kink structure with a linear V-shape around $\nu=\nu_{\rm IRM}$. Here $V_d/E_R=3$, $U_0/E_R=-0.01$, $\Delta_c/E_RN_{\rm at}=-0.155$, $\kappa/E_RN_{\rm at}=0.0075$ and $N=4000$.
}
\label{PhaseDiagram}
 \end{figure}

{\emph {Phenomenological Theory and Phase Diagram}}.--
To characterize the Dicke transitions, in particular around the nesting and IRM fillings, next we derive a phenomenological theory. To begin with, we notice that the DOS presents van Hove singularities and diverges as $|\varepsilon_{\rm ed}-\varepsilon|^{-\frac{1}{2}}$ around the gap edge $\varepsilon_{\rm ed}$, for gaps opened by either Fermi surface nesting effect [Fig.~\ref{Spectrum}(c)] or by IR effect [Fig.~\ref{Spectrum}(d))]. Here, we take $\alpha$ as the real number for further discussion.

For the case of nesting gap, we assume a single-particle dispersion $\varepsilon_k=\tilde{\varepsilon}_{\rm N}-\sqrt{(k^2/2m^*-\tilde{\varepsilon}_{\rm N})^2+4\eta^2\alpha^2}$ for the band below the nesting filling, where $m^*$ is the effective mass around nesting, $\tilde{\varepsilon}_{\rm N}=\varepsilon_{\rm N}-\varepsilon_{n=0,\alpha=0}$ is the Fermi energy at nesting shifted by the band bottom before the superradiant transition ($\alpha = 0$), and $2|\eta\alpha|$ is the gap width. The DOS then reads
\ba
D_\alpha(\varepsilon)=\frac{D_0}{\sqrt{\tilde{\varepsilon}_{\rm N}-\sqrt{\Delta\varepsilon^2-4\eta^2\alpha^2}}}\frac{|\Delta\varepsilon|}{\sqrt{\Delta\varepsilon^2-4\eta^2\alpha^2}},
\ea
where $\Delta\varepsilon=\tilde{\varepsilon}_{\rm N}-\varepsilon$. Notice that this simple form of dispersion can correctly describe the divergence of DOS near the gap edge, and qualitatively capture the overall shape of the actual dispersion down to the single-particle ground state $\varepsilon_{g0}=\tilde{\varepsilon}_{\rm N}-\sqrt{\tilde{\varepsilon}_{\rm N}^2+4\eta^2\alpha^2}$. The prefactor $D_0$ can be solved via the number constraint of total available states $N_0=\int_{\varepsilon_{g0}}^{\varepsilon_{ed}} d\varepsilon D_\alpha(\varepsilon)$, and is quantitatively affected by the details of dispersion. 

To calculate the ground state energy of the system $E_g(\alpha)=E_{\rm vac}(\alpha)+E_{\rm occ}(\alpha)$, we first assume without loss of generality that the shift of the band bottom $E_{\rm vac}(\alpha) \approx \delta \alpha^2$ for small $\alpha$, owing to the $Z_2$ symmetry of the superradiant transition~\cite{Supplementary}. The shift is a joint effect of gap opening and cavity detuning, and acquires a positive prefactor $\delta$ as confirmed by numerical calculation. On the other hand, the occupation energy $E_{\rm occ}(\alpha)=\int_{\varepsilon_{g0}}^\mu d\varepsilon \varepsilon D_\alpha(\varepsilon)$ can be calculated by summing over all states below the chemical potential $\mu$ determined by the filling. By denoting the deviation from the perfect nesting filling $\nu_{\rm N}$ as $\tilde{\nu}_{\rm N} = 1-\nu/\nu_{\rm N}$, we get $\mu \approx \tilde{\varepsilon}_{\rm N} [1-2\sqrt{\tilde{\nu}_{\rm N}^2+(\eta\alpha/\tilde{\varepsilon}_{\rm N})^2}]$ for small $\tilde{\nu}_{\rm N}$, and~\cite{Supplementary}
\ba
E_g(\alpha)\!&=&\!\delta \alpha ^2- N_0\tilde{\varepsilon}_{\rm N}\int_{\tilde{\nu}_{\rm N}}^{1} dx\sqrt{x^2(2-x)^2+\frac{4\eta^2\alpha^2}{\tilde{\varepsilon}_{\rm N}^2}}.
\ea
A phase transition occurs when the overall coefficient of the $\alpha^2$ term changes sign. For a system at nesting filling, the leading order of the integral above is $\alpha^2\log|\alpha|$, hence guarantees a diverging coefficient of the $\alpha^2$ term for small $\alpha$. This observation explains the nesting effect of the Dicke transition, and explains why the critical pumping strength increases rapidly when $\nu$ moves away from $\nu_{\rm N}$.

For a system at the IRM filling $\nu_{\rm IRM}$, the DOS follows a similar form as for the nesting gap
\ba
D_\alpha'(\varepsilon)\! = \!\frac{D_0'}{\sqrt{\tilde{\varepsilon}_{\rm R}\!-\!\sqrt{\Delta\varepsilon'^2-\Delta_{\rm IRM}^2(\alpha)}}}
\frac{|\Delta\varepsilon'|}{\sqrt{\Delta\varepsilon'^2-\Delta_{\rm IRM}^2(\alpha)}},
\ea
with $\tilde{\varepsilon}_{\rm R}=\sqrt{\tilde{\varepsilon}^2_{\rm N}+4\eta^2\alpha^2}-\sqrt{(\tilde{\varepsilon}_{\rm N}-\tilde{\varepsilon}_{\rm IRM})^2+4\eta^2\alpha^2}$, $\Delta\varepsilon'=\tilde{\varepsilon}_{\rm R}-\varepsilon$, $\tilde{\varepsilon}_{\rm IRM}=\varepsilon_{\rm IRM}-\varepsilon_{0,0}$, and $\Delta_{\rm IRM}(\alpha)$ the IRM gap width as a function of $\alpha$. A similar calculation gives~\cite{Supplementary}
\ba
\!\!E_g(\alpha) \!= \!E_{\rm vac}(\alpha) \!\!-\! \!N_0'\tilde{\varepsilon}_{\rm R}\!\!\int_{\tilde{\nu}_{\rm IRM}}^{1}\!\!\!\!\!\!\!dx\sqrt{\!x^2(2-x)^2\!\!+\frac{\Delta_{\rm IRM}^2(\alpha)}{\tilde{\varepsilon}^2_{\rm R}}},
\ea
with $\tilde{\nu}_{\rm IRM} = 1- \nu/\nu_{\rm IRM}$. The key difference of the IRM gap is that $\Delta_{\rm IRM}$ is quadratically dependent on $\alpha$ for small $\alpha$, as can be read from the numerical diagonalization. The superradiant transition can only take place for a finite $\alpha = \alpha_c$, such that the minima of $E_g$ at $\alpha =0$ and $|\alpha_c|$ become degenerate. When the filling is slightly deviated from $\nu_{\rm IRM}$, a perturbation analysis shows that the critical pumping strength of the superradiant transition is linearly dependent on filling, leading to a kink structure~\cite{Supplementary}. The slope of linear dependence is in general different for $\nu < \nu_{\rm IRM}$ and $\nu > \nu_{\rm IRM}$, due to the asymmetry of DOS near $\nu_{\rm IRM}$. The kink structure is a characteristic feature of the IR effect.

The phenomenological theory can be verified by numerical calculation. The ground state energy density ${\cal F}_{\alpha}/E_R N_{\rm at}$ as function of fermion density order $\Theta/N_{\rm at}$ at representative points near the nesting filling and the IRM filling are shown in Figs.~\ref{PhaseDiagram}(a) and \ref{PhaseDiagram}(b), showing typical behavior of second-order and first-order phase transitions, respectively. The phase diagram for different filling factors and critical pumping strengths is given in Fig.~\ref{PhaseDiagram}(c). The normal phase with $\alpha =0$ (bottom) is separated from the superradiant phase (top) by either a second-order transition (dashed line) for large filling, or a first-order transition (solid line) for small filling. The cavity field $|\alpha|$ (false color) is finite at the first-order transition and zero at the second-order transition. For both cases, a finite size scaling is conducted and the results shown here is the thermodynamic limit~\cite{Supplementary}. Three dips with minimal critical pumping are observed at $\nu_{\rm IRM} =1-k_c/k_d$ and $\nu_{\rm N} = k_c/2k_d, 1-k_c/2k_d$. In the inset, we zoom around $\nu=\nu_{\rm IRM}$ to show the kink structure with a linear V-shape. Further, we calculate the inverse participation ratio (IPR) of all occupied states right after the superradiant transition~\cite{Supplementary}, and find that the system is in Anderson localization phase for $\nu<\nu_{\rm AL}$ and quasi-crystal phase for $\nu>\nu_{\rm AL}$. The critical filling $\nu_{\rm AL}$ is marked as a dash-dotted line in Fig.~\ref{PhaseDiagram}(c).

{\emph {Summary}.}--
To summarize, we study the superradiant transition of a 1D Fermi gas coupled to an incommensurate cavity. We find a first-order Dicke transition induced by the indirect resonance effect. The indirect resonance effect can induce superradiant transitions which are highly sensitive to density, with a minimal critical pumping occurring at a specific filling $\nu=\nu_{\rm IRM}$, around which the critical pumping depends linearly on the filling and presents a V-shape kink on the phase diagram. This unique characteristic can act as a smoking-gun evidence for the IR effect. These phenomenons are generally applicable to quasicrystal systems and are expected to be presented in higher-dimensional systems as well. All of our predictions can be tested in future experiments of ultracold Fermi gases in optical cavities.

\emph{Acknowledgements.}--
This work is supported by the National Key R$\&$D Program of China (Grant No. 2018YFA0306501 and 2022YFA1405300), the National Natural Science Foundation of China (Grant No. 11734010, 11774425, 12174358, and 92265208), and the Beijing Natural Science Foundation (Z180013).

\clearpage
\newpage

\setcounter{equation}{0}
\setcounter{figure}{0}
\setcounter{table}{0}
\setcounter{page}{1}
\setcounter{section}{0}
\makeatletter
\makeatother
\global\def\theequation{S\arabic{equation}}
\global\def\thefigure{S\arabic{figure}}%
\global\def\thetable{S\arabic{table}}
\global\def\thepage{S\arabic{page}}
\global\def\thesection{S\arabic{section}}
\renewcommand{\bibnumfmt}[1]{[S#1]}
\renewcommand{\citenumfont}[1]{S#1}

\onecolumngrid
\begin{center}
\large{\bf Supplemental Material for ``Superradiant Transition to a Fermionic Quasicrystal in a Cavity”}  \\
\vspace{0.1in}
\small{Bo-Hao Wu, \it{et. al.} }\\
\end{center}

\section{Diagonalization of Hamiltonian}
\label{sec:diag}

In order to diagonalize the Hamiltonian $\hat{H}_0$ in Eq.~(1) of the main text, we choose the Wannier function of the dipole potential $V_d(x)$ as the basis and adopt the tight-binding approximation. By applying the Bloch theorem to Hamiltonian $\hat{H}_{\rm at}$ and expanding the Bloch wave function $|k\rangle$ in real space by the plane waves, we get
\begin{eqnarray}
\langle x|k\rangle \equiv \psi_k(x)=e^{ikx}u_{k}(x) = e^{ikx}\sum_{l=-l_c}^{+l_c}e^{i2lk_dx}u_{k}(l)\frac{1}{\sqrt{L}} 
= \sum_{l=-l_c}^{+l_c}\frac{1}{\sqrt{L}}e^{i(k+2lk_d)x}u_k(l),
\end{eqnarray}
where $l_c$ is the cutoff of basis, $L=N\pi/k_d$ is the length of the system and $N$ is the number of dipolar lattice site. The quasi-momentum $k\in [-k_d,+k_d)$ falls only in the first Brillouin zone with an interval $\Delta_k=2k_d/N$. The expansion coefficient $u_k(l)$ can be obtained by diagonalizing Hamiltonian $\hat{H}_{\rm at}$ in the plane waves basis. After a shift of zero-point energy by $V_d/2$, the effective Hamiltonian $\hat{H}_{\rm at}(k)$ can be written as
\begin{eqnarray}
\label{Hat}
\hat{H}_{\rm at} (k)= 
\begin{pmatrix}
\frac{(k-2l_ck_d)^2}{2m}&\frac{V_d}{4}&\cdots&0&0\\
\frac{V_d}{4}&\frac{(k-2(l_c-1)k_d)^2}{2m}&\cdots&0&0\\
\vdots&\vdots&\ddots&\cdots&\cdots\\
0&0&\vdots&\frac{(k+2(l_c-1)k_d)^2}{2m}&\frac{V_d}{4}\\
0&0&\vdots&\frac{V_d}{4}&\frac{(k+2l_ck_d)^2}{2m}\\
\end{pmatrix},
\end{eqnarray}
where the matrix elements ${H}^{l,l'}_{\rm at} (k)$ read
\begin{eqnarray}
    {H}^{l,l'}_{\rm at} (k) = \frac{1}{L}\int_{-\frac{L}{2}}^{\frac{L}{2}} 
                            e^{-i(k+2lk_d)x}\hat{H}_{\rm at} e^{i(k+2l'k_d)x} 
                           dx. 
\end{eqnarray}
We then take the single-band approximation and denote the Bloch states of the bottom band as $\psi_k(x)$. Here, we drop the band index to simplify notation. 

The Wannier function $|j\rangle$ in real space can be obtained by Fourier transformation as
\begin{eqnarray}
 \langle x|j\rangle \equiv \phi_{R_{j}}(x)= \frac{1}{\sqrt{N}}\sum_{k}e^{-ikR_{j}}\psi_k(x),
\end{eqnarray}
where $\{R_{j}\}$ is a set of discrete site positions. Further, the fermionic field operator can be expanded by the Wannier basis $\hat{\Psi}(x)=\sum\limits_{j}\phi_{R_{j}}(x)\hat{c}_j$, where $\hat{c}_j$ is the fermionic annihilation operator at position $j$. With that, we obtain the Hamiltonian in the tight-binding approximation
\begin{eqnarray}
\hat{H}=-t\sum_{<i,j>}\hat{c}_{i}^{\dagger}\hat{c}_{j}+\eta(\hat{a}+\hat{a}^{\dagger})\sum_{i}\cos(k_cR_i)\hat{c}_{i}^{\dagger}\hat{c}_{i}
+\frac{U}{2}\hat{a}^{\dagger}\hat{a}\sum_{i}\cos(2k_cR_i)\hat{c}_{i}^{\dagger}\hat{c}_{i}-\delta_c\hat{a}^{\dagger}\hat{a},
\end{eqnarray}
where $\delta_c = \Delta_c-U_0N_{\rm at}/2$ and $N_{\rm at}$ is the atom number.
Here, the nearest-neighbor hopping strength $t$ and the renormalized parameters $\eta$ and $U$ are 
\begin{eqnarray}
t &=& -\int_{-\frac{L}{2}}^{\frac{L}{2}}   \phi_{R_{i+1}}^*(x)\hat{H}_{\rm at}\phi_{R_i}(x)dx,\\
\eta &=& \eta_0\int_{-\frac{L}{2}}^{\frac{L}{2}}  |\phi_{R_{i}}(x)|^2 \cos[k_c(x-R_i)]dx,\\
U &=& U_0\int_{-\frac{L}{2}}^{\frac{L}{2}}  |\phi_{R_{i}}(x)|^2 \cos[2k_c(x-R_i)]dx.
\end{eqnarray}
In Tab. S1, we list the numerical results of $t$, $\eta$, and $U$ for some typical choices of lattice depth $V_d/E_R$ with $E_R$ the recoil energy.
\begin{table}[tbp]
\begin{center}
\label{Table1}
\caption{Numerical results of $t$, $\eta$, and $U$ for different lattice depths.}
\begin{tabular}{|c|c|c|c|}
\hline \textbf{$V_d/E_R$} & \textbf{$4t/E_R$} & \textbf{$\eta/\eta_0$} & \textbf{$U/U_0$}\\
\hline  3&0.444109&0.909491&0.855470 \\
\hline  5&0.263069&0.938906&0.892875 \\
\hline  10&0.076730&0.962952&0.930528\\
\hline  15&0.026075&0.971324&0.945272\\
\hline  20&0.009965&0.975816&0.953465\\
\hline
\end{tabular}
\end{center}
\end{table}

Finally, we employ the mean field approximation for the cavity mode $\langle\hat{a}\rangle = \alpha$ and write down the Hamiltonian
\begin{eqnarray}
    \label{Ham_alpha}
    \hat{H}(\alpha) =-t\sum_{\langle i ,j\rangle}\hat{c}_{i}^{\dagger}\hat{c}_{j}+\eta(\alpha+\alpha^*)\sum_{i}\cos(k_cR_i)\hat{c}_{i}^{\dagger}\hat{c}_{i}
    +\frac{U}{2}\alpha^*\alpha\sum_{i}\cos(2k_cR_i)\hat{c}_{i}^{\dagger}\hat{c}_{i}-\delta_c\alpha^*\alpha,
\end{eqnarray}
from which the single-particle spectrum $\{\varepsilon_{n,\alpha}\}$ and the corresponding eigenstates $\{ |\phi_{n,\alpha}\rangle\}$ can be calculated by diagonalizing Eq.~(\ref{Ham_alpha}) under open boundary condition. 

\begin{figure}[tbp]
    \hspace{-2ex}\includegraphics[width=8.5cm]{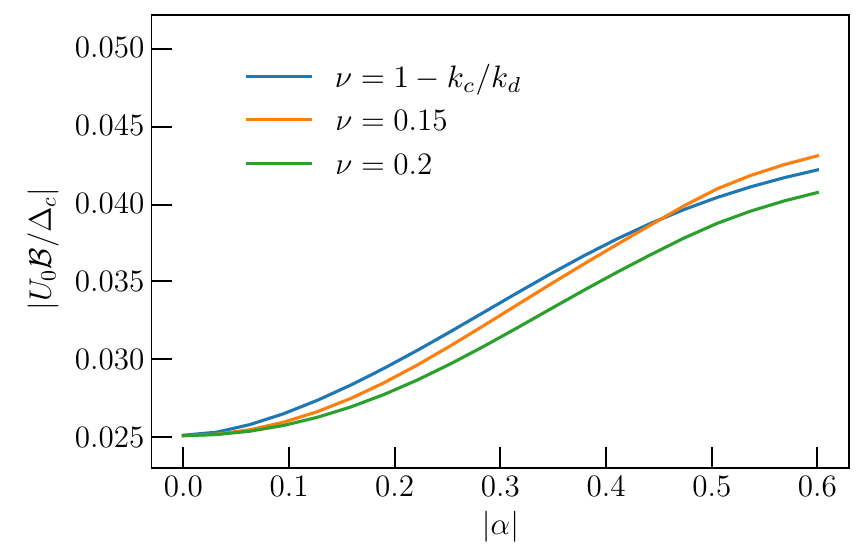}
    \caption{Numerical results of $|U_0{\cal B}/\Delta_c|$ as a function of cavity mode $|\alpha|$ at different filling $\nu$. Here, we use $V_d/E_R=3$, $\eta_0/E_R=-0.273$, $U_0/E_R=-0.01$, $\Delta_c/E_RN_{\rm at}=-0.155$, $\kappa/E_RN_{at}=0.0075$ and system size $N=4000$.}
       \label{UB}
\end{figure}

Before concluding this section, we stress that in the main text a fixed phase of the cavity mode $\alpha$ is assumed for simplicity, which is valid under the condition $\Delta_c\gg U_0{\cal B}$. To verify this assumption, we plot in Fig.~\ref{UB} the ratio $|U_0{\cal B}/\Delta_c|$. This quantity remains to be less than 5\% for different filling when $|\alpha| < 0.6$, validating the assumption of a fixed phase in all parameter regions discussed in this work.

\section{Finite size analysis}
\label{sec:finite}

\subsection{$\mathcal{Z}_2$ symmetry of cavity mode}
\begin{figure}[h]
    \hspace{-2ex}\includegraphics[width=13cm]{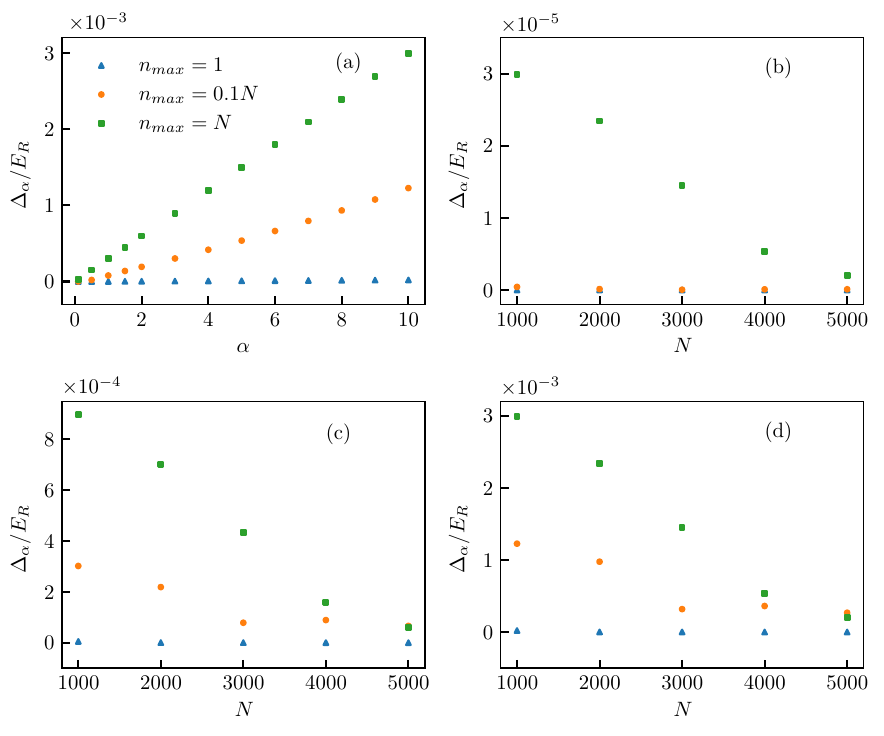}
    \caption{(a) The energy imbalance $\Delta_{\alpha}/E_R$ as a function of $\alpha$ for different $n_{\rm max}$ with $N = 1000$. (b, c, d) The energy imbalance $\Delta_{\alpha}$ as a function of system size $N$ for $\alpha = 0.1, 3, 10$, respectively. Parameters used here are $V_d/E_R=3$, $\eta_0/E_R=-0.273$ and $U_0/E_R=-0.01$, and $\alpha$ is assumed to be real.} 
\label{Z2}
\end{figure}

Owing to the incommensurability of $k_c/k_d$, one in general cannot assume {\it a priori} a $\mathcal{Z}_2$ symmetry of the cavity mode, in particular for a finite-size system. To clarify this issue, we define the imbalance of energy at $\alpha$ and $-\alpha$ as
\begin{eqnarray}
    \frac{\Delta_{\alpha}}{E_{R}} = \frac{1}{E_{R}n_{\rm max}}
                            \sum_{n=1}^{n_{\rm max}} |\varepsilon_n(\alpha) - \varepsilon_n(-\alpha) |,
\end{eqnarray}
where $n_{\rm max}$ is the total number of energy levels under consideration. In Fig.~\ref{Z2}, we show the numerical result of $\Delta_{\alpha}/E_R$ by varying $\alpha$ and $N$. Indeed, we find a nonzero $\Delta_\alpha/E_R$ for a finite-size system with $\alpha >0$. However, the energy imbalance is quite small for a large enough system with $\alpha \lesssim 1$, which is the case of particular interest in this work.
%

\subsection{Phase diagram}
\begin{figure}[tbp]
    \hspace{-2ex}\includegraphics[width=16.5cm]{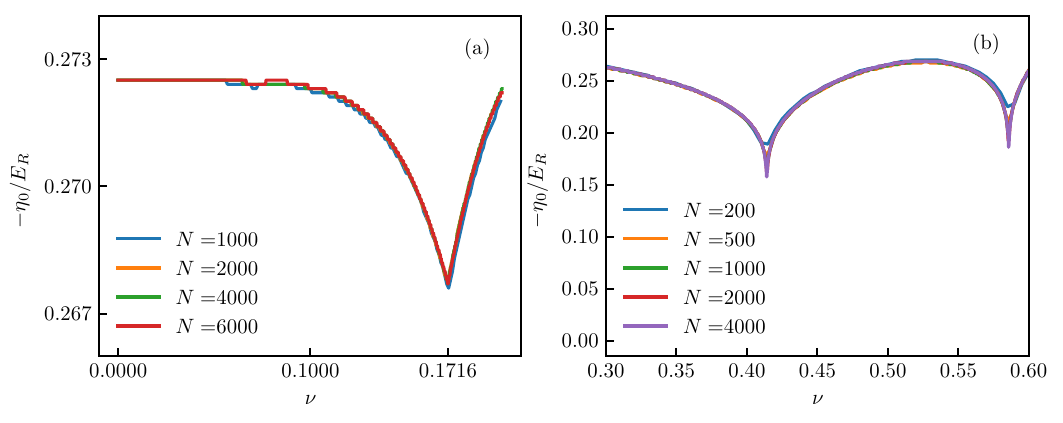}
    \caption{The phase diagram around (a) the IRM filling and (b) the nesting filling for different size $N$. Parameters used here are $V_d/E_R=3$, $U_0/E_R=-0.01$, $\Delta_c/E_RN_{\rm at}=-0.155$ and $\kappa/E_RN_{at}=0.0075$.}\label{PD1}
\end{figure}

\begin{figure}[tbp]
    \hspace{-2ex}\includegraphics[width=8.8cm]{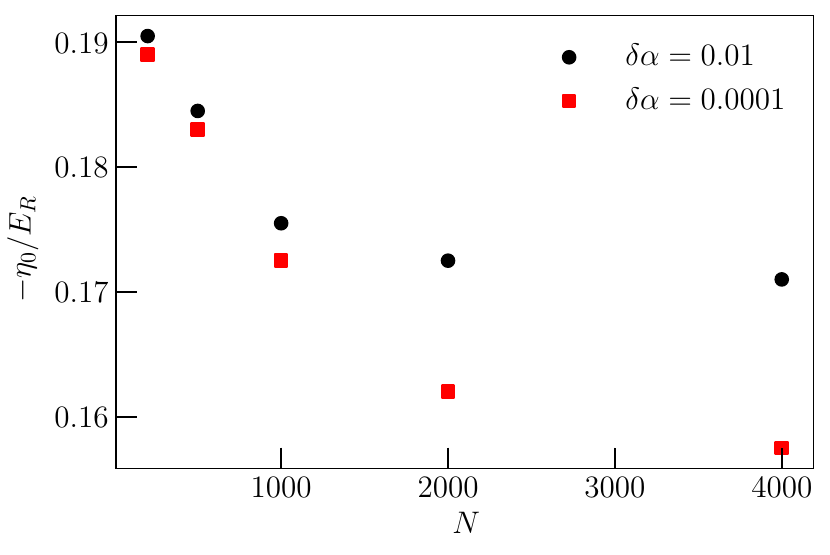}
    \caption{The transition point at the nesting filling $\nu=k_c/2k_d$ for different system size $N$. The black circles and the red squares denote results of different precisions of $\alpha$. Parameters used here are $V_d/E_R=3$, $U_0/E_R=-0.01$, $\Delta_c/E_RN_{\rm at}=-0.155$ and $\kappa/E_RN_{at}=0.0075$.}
    \label{NESTING}
\end{figure}

In this section, we provide the finite size scaling of the phase diagram. Firstly, we analyze the dip around the indirect resonance modified (IRM) filling $\nu_{\rm IRM}=1-k_c/k_d$, at which a first-order Dicke transition takes place. As shown in Fig.~\ref{PD1}(a), the phase boundary for systems of different size converge and the dip has a finite critical pumping strength $\eta_0/E_R \approx 0.268$. For systems with nesting filling, i.e., $\nu=k_c/2k_d$ and $1-k_c/2k_d$, the superradiant transition is of second order. As shown in Fig.~\ref{PD1}(b), the phase boundary for systems of different size converge almost everywhere except at the nesting filling. By increasing the precision of $\alpha$ and the size $N$, we find that the critical pumping strength tends to zero as shown in Fig.~\ref{NESTING}. 
%

\section{Inverse participation ratio}
\label{sec:IPR}

For a given eigenstate $\varphi_n(\alpha)$, we can define the single state inverse participation ratio (IPR) as
\begin{eqnarray}\label{ipr}
IPR^{(n)}(\alpha) = \frac{ \sum_{j}|\langle j| \varphi_n (\alpha)\rangle|^4 }
{ (\sum_{j}|\langle j| \varphi_n (\alpha)\rangle|^2)^2 }.
\end{eqnarray}
For a spatially extended state, $IPR^{(n)}(\alpha)$ tends to zero, while for a localized state it approaches unity. In Fig.~2(a) of the main text, we show the result of $IPR^{(n)}(\alpha)$ for all single-particle eigenstates in false color. 

\begin{figure}[tbp]
    \hspace{-2ex}\includegraphics[width=16cm]{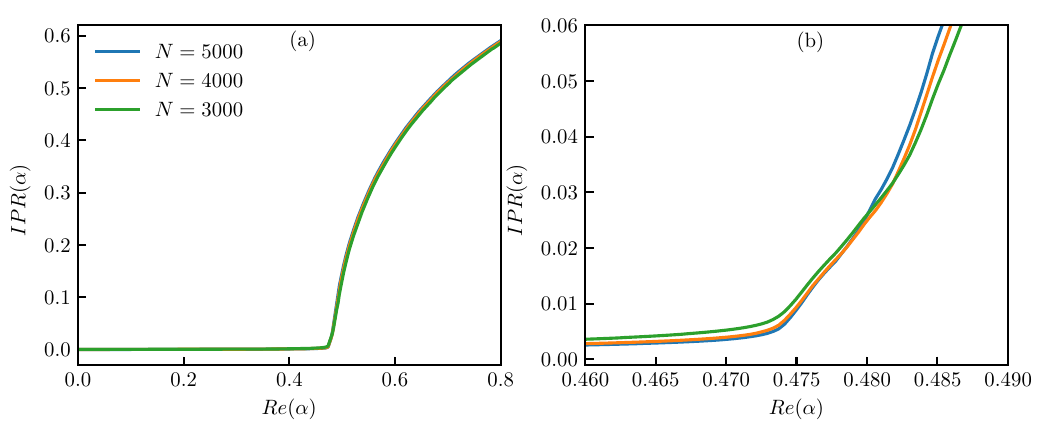}
    \caption{(a) The mean inverse participation ratio $IPR (\alpha)$ as a function of $Re(\alpha)$ for different system size $N$. (b) A zoom-in of (a) showing the intersection of curves. The parameters are $V_d/E_R=3$, $\eta_0/E_R=-0.273$, $U_0/E_R=-0.01$, $\Delta_c/E_RN_{\rm at}=-0.155$ and $\kappa/E_RN_{at}=0.0075$.}\label{IPR}
\end{figure}

By averaging over all eigenstates, we can define the mean inverse participation ratio for the total spectrum 
\begin{eqnarray}\label{mipr}
    IPR (\alpha)= \frac{1}{N}\sum_n IPR^{(n)}(\alpha),
\end{eqnarray}
which can be used to distinguish localized phase and extended phase. Specifically, for a given pumping strength $\eta_0$, we calculate $IPR$ as a function of $\alpha$ for systems with different size, as shown in Fig.~\ref{IPR}. The critical $\alpha$ of the localized--extended phase transition is determined by the average of intersection of curves for different sizes. Then we vary $\eta_0$ and choose the largest $\alpha$ to plot the Anderson localization transition line (dashed--dotted) in Fig.~3(c) of the main text.
%

\section{Phenomenological Theory}
\label{sec:eff}

In this section, we will derive a phenomenological density of states (DOS) and calculate the effective ground state energy density
\ba
E_g(\alpha)=E_{\rm vac}(\alpha)+E_{\rm occ}(\alpha),
\ea
where $E_{\rm vac}(\alpha)$ is the single particle ground state energy shift, or vacuum energy in brief, and $E_{\rm occ}(\alpha)$ is the occupation energy by filling the states below the Fermi level. The occupation energy can be written as
\ba
E_{\rm occ}(\alpha)=\int d\varepsilon D_\alpha(\varepsilon)\varepsilon,
\ea
with $D_\alpha(\varepsilon)$ the DOS. In the following discussion, we take $\kappa=0$ and assume $\alpha$ is real for simplicity.

\subsection{Direct resonance at nesting filling}

We first  study the direct resonance at nesting filling when $k_F=k_c/2$. To describe the van Hove singularity at the band edge, we assume an approximate dispersion
\ba
\varepsilon=\tilde{\varepsilon}_{\rm N}-\sqrt{\left(\frac{k^2}{2m^*}-\tilde{\varepsilon}_{\rm N}\right)^2+4\eta^2\alpha^2}
\ea
for the bottom band below nesting energy $\tilde{\varepsilon}_{\rm N}=\varepsilon_{k_c/2}$, where $m^*$ is the effective mass considering the lattice correction, and $2|\eta\alpha|$ is the resonance induced gap. We stress that this simple form of dispersion can correctly capture the diverging characteristics at band edges, which are crucial to determine the critical behavior of the superradiant phase transition around the resonance point. The DOS can then be written as
\ba
D_\alpha(\varepsilon)=\frac{D_0}{\sqrt{\tilde{\varepsilon}_{\rm N}-\sqrt{\Delta\varepsilon^2-4\eta^2\alpha^2}}}\frac{|\Delta\varepsilon|}{\sqrt{\Delta\varepsilon^2-4\eta^2\alpha^2}},
\ea
where $\Delta\varepsilon \equiv \varepsilon-\tilde{\varepsilon}_{\rm N}$, and $D_0$ is a constant determined by the constraint of particle number $N_0=\int d\varepsilon D_\alpha(\varepsilon)$. Indeed, we can easily find that
\ba
N_0&=&\int_{\tilde{\varepsilon}_{\rm N}-\sqrt{\tilde{\varepsilon}_{\rm N}^2+4\eta^2\alpha^2}}^{\tilde{\varepsilon}_{\rm N}-2\eta\alpha}d\varepsilon \frac{D_0}{\sqrt{\tilde{\varepsilon}_{\rm N}-\sqrt{\Delta\varepsilon^2-4\eta^2\alpha^2}}}\frac{|\Delta\varepsilon|}{\sqrt{\Delta\varepsilon^2-4\eta^2\alpha^2}}=D_0\sqrt{2\eta\alpha}\int_{1}^{\frac{\sqrt{\tilde{\varepsilon}_{\rm N}^2+4\eta^2\alpha^2}}{2\eta\alpha}} \!\!\!\!\!\!dx \frac{1}{\sqrt{\frac{\tilde{\varepsilon}_{\rm N}}{2\eta\alpha}-\sqrt{x^2-1}}}\frac{x}{\sqrt{x^2-1}}\nonumber\\
&=&D_0\sqrt{2\eta\alpha}\int_0^{\arccos\frac{2\eta\alpha}{\sqrt{\tilde{\varepsilon}_{\rm N}^2+4\eta^2\alpha^2}}} \frac{d\theta}{\cos^2\theta}\frac{1}{\sqrt{\frac{\tilde{\varepsilon}_{\rm N}}{2\eta\alpha}-\tan\theta}}=D_0\sqrt{2\eta\alpha}\int_0^{\frac{\tilde{\varepsilon}_{\rm N}}{2\eta\alpha}} \frac{dy}{\sqrt{\frac{\tilde{\varepsilon}_{\rm N}}{2\eta\alpha}-y}}=2D_0\sqrt{\tilde{\varepsilon}_{\rm N}},
\ea
i.e., $D_0=N_0/2\sqrt{\tilde{\varepsilon}_{\rm N}}$ is a constant independent of $\alpha$.

For a system with a filling $\nu$ slightly lower than the nesting filling $\nu_{\rm N}$, the number constraint reads
\ba
\frac{N_0 \nu}{\nu_{\rm N}} &=&\int_{\tilde{\varepsilon}_{\rm N}-\sqrt{\tilde{\varepsilon}_{\rm N}^2+4\eta^2\alpha^2}}^{\mu}d\varepsilon D_\alpha(\varepsilon) = 
D_0\sqrt{2\eta\alpha}\int_{\frac{\sqrt{(\tilde{\varepsilon}_{\rm N}-\mu)^2-4\eta^2\alpha^2}}{2\eta\alpha}}^{\frac{\tilde{\varepsilon}_{\rm N}}{2\eta\alpha}}\frac{dy}{\sqrt{\frac{\tilde{\varepsilon}_{\rm N}}{2\eta\alpha}-y}},
\ea
from which we can obtain the chemical potential
\ba
\mu = \tilde{\varepsilon}_{\rm N} - \sqrt{\left[ 1- \left(\frac{\nu}{\nu_{\rm N}}\right)^2 \right]^2\tilde{\varepsilon}_{\rm N}^2 + 4\eta^2\alpha^2}.
\label{eq:mu1}
\ea
The occupation energy then becomes
\ba
E_{\rm occ}(\alpha)&=&\int_{\tilde{\varepsilon}_{\rm N}-\sqrt{\tilde{\varepsilon}_{\rm N}^2+4\eta^2\alpha^2}}^{\mu} d\varepsilon \varepsilon D_\alpha(\varepsilon)=\int_{\tilde{\varepsilon}_{\rm N}-\mu}^{\sqrt{\tilde{\varepsilon}_{\rm N}^2+4\eta^2\alpha^2}}d\varepsilon(\tilde{\varepsilon}_{\rm N}-\varepsilon)D_\alpha(\tilde{\varepsilon}_{\rm N}-\varepsilon).
\ea
Under the condition $1-\nu/\nu_{\rm N} \to 0$, it is easy to obtain
\ba
N_0\tilde{\varepsilon}_{\rm N}-E_{\rm occ}(\alpha)&=&\int_{\tilde{\varepsilon}_{\rm N}-\mu}^{\sqrt{\tilde{\varepsilon}_{\rm N}^2+4\eta^2\alpha^2}}d\varepsilon \frac{D_0}{\sqrt{\tilde{\varepsilon}_{\rm N}-\sqrt{\varepsilon^2-4\eta^2\alpha^2}}}\frac{\varepsilon^2}{\sqrt{\varepsilon^2-4\eta^2\alpha^2}}\nonumber\\
&=&D_0(2\eta\alpha)^{3/2}\int_{\frac{\tilde{\varepsilon}_{\rm N}-\mu}{2\eta\alpha}}^{\frac{\sqrt{\tilde{\varepsilon}_{\rm N}^2+4\eta^2\alpha^2}}{2\eta\alpha}}  \frac{dx}{\sqrt{\frac{\tilde{\varepsilon}_{\rm N}}{2\eta\alpha}-\sqrt{x^2-1}}}\frac{x^2}{\sqrt{x^2-1}}\nonumber\\
&=&D_0(2\eta\alpha)^{3/2}\int_{\frac{\sqrt{(\tilde{\varepsilon}_{\rm N}-\mu)^2-4\eta^2\alpha^2}}{2\eta\alpha}}^{\frac{\tilde{\varepsilon}_{\rm N}}{2\eta\alpha}} dy\frac{\sqrt{1+y^2}}{\sqrt{\frac{\tilde{\varepsilon}_{\rm N}}{2\eta\alpha}-y}}.
\ea
By defining $y \equiv \frac{\varepsilon_{\rm N}}{2\eta\alpha}\sin^2\phi$, we have
\ba
N_0\tilde{\varepsilon}_{\rm N}-E_{\rm occ}(\alpha)&=&N_0\tilde{\varepsilon}_{\rm N}\int_{\arcsin\sqrt{\frac{\sqrt{(\tilde{\varepsilon}_{\rm N}-\mu)^2-4\eta^2\alpha^2}}{\tilde{\varepsilon}_{\rm N}}}}^{\frac{\pi}{2}}d\phi \sin\phi \sqrt{\sin^4\phi+\frac{4\eta^2\alpha^2}{\tilde{\varepsilon}_{\rm N}^2}}\nonumber\\
&\approx&N_0\tilde{\varepsilon}_{\rm N}\int_{\sqrt{2(1-\nu/\nu_{\rm N}})}^{\frac{\pi}{2}}d\phi \sin\phi \sqrt{\sin^4\phi+\frac{4\eta^2\alpha^2}{\tilde{\varepsilon}_{\rm N}^2}},
\ea
where the expression of chemical potential Eq.~(\ref{eq:mu1}) is substituted. Finally, we obtain
\ba
E_{\rm occ}(\alpha)&=&N_0\tilde{\varepsilon}_{\rm N}-N_0\tilde{\varepsilon}_{\rm N}\int_{\sqrt{2(1-\nu/\nu_{\rm N})}}^{\frac{\pi}{2}}d\phi \sin\phi \sqrt{\sin^4\phi+\frac{4\eta^2\alpha^2}{\tilde{\varepsilon}_{\rm N}^2}}
\nonumber\\
&\approx&N_0\tilde{\varepsilon}_{\rm N}-N_0\tilde{\varepsilon}_{\rm N}\int_{\tilde{\nu}_{\rm N}}^1 dx \sqrt{x^2(2-x)^2+\frac{4\eta^2\alpha^2}{\tilde{\varepsilon}_{\rm N}^2}}
\label{eq:Eocc}
\ea
with $\tilde{\nu}_{\rm N} \equiv 1- \nu/\nu_{\rm N}$.
Notice that the first term on the right-hand-side is a constant and can be dropped out as a zero-point energy. Then we get the expression used in Eq.~(8) of the main text.

For a system with filling slightly higher than the nesting filling, an analogous derivation leads to a similar result as Eq.~(\ref{eq:Eocc}) with the lower bound of the integral changing to $\nu/\nu_{\rm N} -1$.

\subsection{Indirect resonance modulation at IRM filling}

Following a similar derivation as in the previous section, we can get the ground state energy $E_g$ around the IRM filling $\nu_{\rm IRM}$ analogous to Eq.~(\ref{eq:Eocc})
\ba
E_g(\alpha)&=& E_{\rm vac}(\alpha) - N_0'\tilde{\varepsilon}_{\rm R}\int_{\tilde{\nu}_{\rm IRM}}^{1}dx  \sqrt{{x^2(2-x)^2}+\frac{\Delta_{\rm IRM}^2(\alpha)}{\tilde{\varepsilon}^2_{\rm R}}},
\ea
with $E_{\rm vac}(\alpha)=(1-\tilde{\nu}_{\rm IRM})N_0'\varepsilon_{0,\alpha}-\Delta_c\alpha^2$, 
$\tilde{\varepsilon}_{\rm R}=\sqrt{\tilde{\varepsilon}^2_{\rm N}+4\eta^2\alpha^2}-\sqrt{(\tilde{\varepsilon}_{\rm N}-\tilde{\varepsilon}_{\rm IRM})^2+4\eta^2\alpha^2}$, 
and $\tilde{\nu}_{\rm IRM} \equiv 1- \nu/\nu_{\rm IRM}$.
The gap of the single-particle dispersion $\Delta_{\rm IRM}$ depends on $\alpha$ quadratically for small $\alpha$. This behavior is in stark contrast to the case of nesting filling where a linear dependence is found. The prefactor of the $\alpha^2$ term in $E_g$ remains positive when $\alpha$ is small, and the superradiant transition can only take place at a finite $\alpha$.

\begin{figure}[tbp]
\hspace{-2ex} \includegraphics[width=8.8cm]{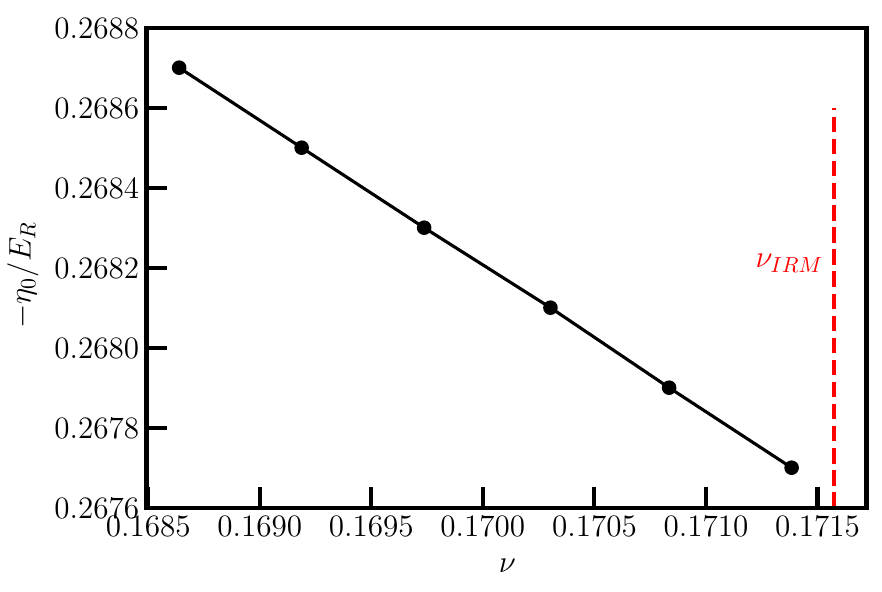}
\caption{The critical pumping $\eta_0/E_R$ as a function of filling $\nu$ near the IRM filling. Parameters used are $V_d/E_R=3$, $U_0/E_R=-0.01$, $\Delta_c/E_RN_{\rm at}=-0.1774$, and $N=4000$.}
\label{fig:IRphase}
\end{figure}

To analyze the superradiant transition around the IRM filling, we denote the critical pumping strength as $\eta_0$ and the critical cavity field as $\alpha_{c}$ for a specific value of $\tilde{\nu}_{\rm IRM} = \tilde \nu_0$. Since the phase transition is of first order, we have
\ba
E_{g}(\tilde{\nu}_0,\alpha=0,\eta_0) = E_{g}(\tilde{\nu}_0,\alpha=\alpha_{c},\eta_0).
\ea
Next we consider an increment of filling $\tilde{\nu}_{\rm IRM} \to \tilde \nu_0 + \delta \tilde \nu$, and denote the corresponding critical pumping strength as $\eta_0 + \delta \eta$ and critical cavity field as $\alpha_{c} + \delta \alpha$. The condition of phase transition then reads
\ba
E_{g}(\tilde{\nu}_0+\delta \tilde \nu,\alpha=\delta\alpha, \eta_0+\delta \eta) = E_{g}(\tilde{\nu}_0+\delta \tilde \nu,\alpha=\alpha_{c}+\delta\alpha,\eta_0+\delta \eta).
\ea
Expanding the expression above to linear order of $\delta \tilde\nu$, $\delta \eta$ and $\delta \alpha$, we can easily conclude that the critical pumping field is linearly dependent on filling with
\ba \label{result}
\frac{\delta \eta}{\delta \tilde \nu} &=& \frac{\zeta_1\zeta_6-\zeta_3\zeta_4}{\zeta_3\zeta_5-\zeta_2\zeta_6},
\ea
where
\ba
\zeta_1 &=& \frac{1}{N_0'}\frac{\partial\left[E_{\rm vac}(\tilde{\nu},\alpha=0,\eta_0)- E_{\rm vac}(\tilde{\nu},\alpha=\alpha_c,\eta_0)\right]}{\partial \tilde{\nu}}\bigg|_{\tilde{\nu}=\tilde{\nu}_0}+\left[ T(\tilde{\nu}_0,\alpha=0,\eta_0)-T(\tilde{\nu}_0,\alpha=\alpha,\eta_0)\right],
\nonumber \\
\zeta_2 &=& \frac{1}{N_0'}\frac{\partial\left[E_{\rm vac}(\tilde{\nu}_0,\alpha=0,\eta)- E_{\rm vac}(\tilde{\nu}_0,\alpha=\alpha_c,\eta)\right]}{\partial \eta}\bigg|_{\eta=\eta_0}-\int_{\tilde{\nu}_0}^{1}dx \frac{\partial \left[ T(x,\alpha=0,\eta)-T(x,\alpha=\alpha_c,\eta)\right]}{\partial \eta}\bigg|_{\eta=\eta_0},
\nonumber \\
\zeta_3 &=& \frac{1}{N_0'}\left[\frac{\partial E_{\rm vac}(\tilde{\nu}_0,\alpha,\eta_0)}{\partial \alpha}\bigg|_{\alpha=\alpha_c}-\frac{\partial E_{\rm vac}(\tilde{\nu}_0,\alpha,\eta_0)}{\partial \alpha}\bigg|_{\alpha=0}\right]-\int_{\tilde{\nu}_0}^{1}dx \left[ \frac{\partial T(x,\alpha,\eta_0)}{\partial\alpha}\bigg|_{\alpha=\alpha_c}-\frac{\partial T(x,\alpha,\eta_0)}{\partial\alpha}\bigg|_{\alpha=0}\right],
\nonumber \\
\zeta_4 &=& \frac{1}{N_0'}\frac{\partial^2 E_{\rm vac}(\tilde{\nu},\alpha,\eta_0)}{\partial \alpha\partial\tilde{\nu}}\bigg|_{\tilde{\nu}=\tilde{\nu}_0,\alpha=\alpha_c}+\frac{\partial T(\tilde{\nu}_0,\alpha,\eta_0)}{\partial\alpha}\bigg|_{\alpha=\alpha_c},
\nonumber \\
\zeta_5 &=& \frac{1}{N_0'}\frac{\partial^2 E_{\rm vac}(\tilde{\nu}_0,\alpha,\eta)}{\partial\alpha\partial\eta}\bigg|_{\eta=\eta_0,\alpha=\alpha_c}-\int_{\tilde{\nu}_0}^{1}dx \frac{\partial^2T(x,\alpha,\eta)}{\partial\alpha\partial\eta}\bigg|_{\eta=\eta_0,\alpha=\alpha_c},
\nonumber \\
\zeta_6 &=& -\frac{1}{N_0'}\frac{\partial^2 E_{\rm vac}(\tilde{\nu}_0,\alpha,\eta_0)}{\partial\alpha^2}\bigg|_{\alpha=\alpha_c}+\int_{\tilde{\nu}_0}^{1}dx \frac{\partial^2 T(x,\alpha,\eta_0)}{\partial\alpha^2}\bigg|_{\alpha=\alpha_c}.
\ea
and
\ba
T(x,\alpha,\eta) &=& \sqrt{x^2(2-x)^2+\frac{\Delta_{\rm IRM}^2(\alpha)}{\tilde{\varepsilon}_{\rm R}^2}}.
\ea
The linear dependence Eq.~(\ref{result}) of the critical pumping strength on filling can be numerically verified, as shown in Fig.~\ref{fig:IRphase}.


\begin{thebibliography}{2}

\bibitem{Esslinger_2007}
F. Brennecke, T. Donner, S. Ritter, T. Bourdel, M. K{\"o}hl, and T. Esslinger, Cavity QED with a Bose-Einstein condensate, 
{\emph{Nature (London)}} {\bf450}, 268 (2007).

\bibitem{Esslinger_2013}
H. Ritsch, P. Domokos, F. Brennecke, and T. Esslinger, Cold atoms in cavity-generated dynamical optical potentials, 
{\emph{Rev. Mod. Phys.}} {\bf85}, 553 (2013).

\bibitem{Esslinger_2010}
K. Baumann, C. Guerlin, F. Brennecke, and T. Esslinger, Dicke quantum phase transition with a superfluid gas in an optical cavity, 
{\emph{Nature (London)}} {\bf 464}, 1301 (2010).

\bibitem{Esslinger_2011}
K. Baumann, R. Mottl, F. Brennecke, and T. Esslinger, Exploring Symmetry Breaking at the Dicke Quantum Phase Transition, 
 {\emph{Phys. Rev. Lett.}} {\bf107}, 140402 (2011).

\bibitem{Esslinger_2012}
R. Mottl, F. Brennecke, K. Baumann, R. Landig, T. Donner, and T. Esslinger, Roton-Type Mode Softening in a Quantum Gas with Cavity-Mediated Long-Range Interactions, 
{\emph{Science}} {\bf 336}, 1570 (2012).

\bibitem{Esslinger_2015}
R. Landig, F. Brennecke, R. Mottl, T. Donner, and T. Esslinger, Measuring the dynamic structure factor of a quantum gas undergoing a structural phase transition, 
{\emph{Nat. Commun.}} {\bf 6}, 7046 (2015).

\bibitem{Simons_2014}
J. Keeling, M. J. Bhaseen, and B. D. Simons, Fermionic Superradiance in a Transversely Pumped Optical Cavity, 
{\emph{Phys. Rev. Lett.}} {\bf112}, 143002 (2014).

\bibitem{Piazza_2014}
F. Piazza and P. Strack, Umklapp Superradiance with a Collisionless Quantum Degenerate Fermi Gas, 
{\emph{Phys. Rev. Lett.}} {\bf112}, 143003 (2014).

\bibitem{Chen_2014}
Y. Chen, Z. Yu, and H. Zhai, Superradiance of Degenerate Fermi Gases in a Cavity, 
{\emph{Phys. Rev. Lett.}} {\bf112}, 143004 (2014).

\bibitem{Wu_2021}
X. Zhang, Y. Chen, Z. Wu, J. Wang, J. Fan, S. Deng, and H. Wu, Observation of a superradiant quantum phase transition in an intracavity degenerate Fermi gas, 
{\emph{Science}} {\bf373}, 1359 (2021).

\bibitem{Kjaegaard_2021}
A. B. Deb and N. Kj\ae rgaard, Observation of Pauli blocking in light scattering from quantum degenerate fermions, 
{\emph{Science}} {\bf 374}, 972 (2021).

\bibitem{Ketterle_2021}
Y. Margalit, Y.-K. Lu, F. \c{C}. Top, and W. Ketterle, Pauli blocking of light scattering in degenerate fermions, 
{\emph{Science}} {\bf 374}, 976 (2021).

\bibitem{Ye_2021}
C. Sanner, L. Sonderhouse, R. B. Hutson, L. Yan, W. R. Milner, and J. Ye, Pauli blocking of atom-light scattering, 
{\emph{Science}} {\bf 374}, 979 (2021).

\bibitem{Zheng_Cooper}
W. Zheng and N. R. Cooper, Anomalous diffusion in a dynamical optical lattice, 
{\emph{Phys. Rev. A}} {\bf97}, 021601(R) (2018).

\bibitem{Sun_2020}
H. Yin, J. Hu, A.-C. Ji, G. Juzeli{\=u}nas, X.-J. Liu, and Q. Sun, Localization Driven Superradiant Instability, 
{\emph{Phys. Rev. Lett.}} {\bf124}, 113601 (2020).

\bibitem{Piazza_2019}
F. Mivehvar, H. Ritsch, and F. Piazza, Emergent Quasicrystalline Symmetry in Light-Induced Quantum Phase Transitions, 
{\emph{Phys. Rev. Lett.}} {\bf123}, 210604 (2019).

\bibitem{Mott56}
N. F. Mott, On the Transition to Metallic Conduction in Semiconductors, 
{\emph{Can. J. Phys.}} {\bf 34}, 1356 (1956).

\bibitem{Mott69}
N. F. Mott, Conduction in non-crystalline materials, 
{\emph{Philos. Mag.}} {\bf 19}, 835 (1969).

\bibitem{Mott79}
N. F. Mott and E. A. Davis, {\it Electronic Processes in Non-Crystalline Materials.} 
(Oxford University Press, New York, 1979)

\bibitem{Piazza_2014_2}
F. Piazza and P. Strack, Quantum kinetics of ultracold fermions coupled to an optical resonator, 
{\emph{Phys. Rev. A}} {\bf90}, 043823 (2014).

\bibitem{Supplementary} 
See Supplemental Material for detailed information on the diagonalization of Hamiltonian, finite size analysis, calculation of inverse participation ratio, and the derivation of the phenomenological theory.

\end{thebibliography}
\end{document}